\documentclass[twocolumn]{aastex631}

\usepackage{mhchem}
\usepackage{threeparttable}
\usepackage{xcolor}
\usepackage[T1]{fontenc}

\newcommand*\msun{\mathrm{M}_\odot}

\newcommand*\mearth{\mathrm{M}_\oplus}

\newcommand*\OmegaK{\Omega_\mathrm{K}}

\begin{document}

\title{Saturn's Formation at the Carbon Dioxide Iceline}

\correspondingauthor{Artyom Aguichine}
\email{aaguichi@astro.unam.mx}

\author[0000-0002-8949-5956]{Artyom Aguichine}
\affiliation{Department of Astronomy and Astrophysics, University of California, Santa Cruz, CA, USA}
\affiliation{Instituto de Astronomía, Universidad Nacional Autónoma de México, Apartado Postal 106, CP 22800 Ensenada, Baja California, México}

\author[0000-0001-5323-6453]{Olivier Mousis}
\affiliation{Solar System Science and Exploration Division, Southwest Research Institute, 1301 Walnut St, Ste 400, Boulder, CO, USA}

\author[0000-0002-2342-3458]{Robin M. Canup}
\affiliation{Solar System Science and Exploration Division, Southwest Research Institute, 1301 Walnut St, Ste 400, Boulder, CO, USA}




\begin{abstract}
{Saturn's bulk metallicity and atmospheric volatile enrichments exceed those of Jupiter. Combined with their similar heavy-element masses, this suggests that Saturn formed farther out in a protoplanetary disk whose metallicity increased with heliocentric distance.} Numerous models argue that Jupiter formed near the H$_2$O iceline; Saturn’s composition is therefore expected to reflect accretion beyond this region. We test whether Saturn’s elemental enrichments can be reproduced by formation near the CO$_2$ or CO icelines using a one-dimensional viscous $\alpha$-disk model that tracks the transport, sublimation, and recondensation of 13 volatile carriers. We compute time-dependent radial distributions of solids and vapors for $\alpha = 5\times10^{-4}$–$5\times10^{-3}$ and varied CO/CO$_2$ abundance ratios, and compare the resulting metallicities and enrichments with Saturn’s observed composition. Strong carbon and oxygen enrichment peaks result from the cold-finger effect, most prominently at the CO$_2$ iceline, where supersolar metallicities comparable to Saturn’s envelope are achieved only for $\alpha \le 10^{-3}$. Smaller C and O enrichment peaks also occur at the CO iceline, but the local protosolar nebula metallicity there never reaches Saturn’s inferred value. At $\sim$0.2--0.3 Myr, carbon and phosphorus enrichments in this region match Saturn's atmospheric values, while noble-gas abundances remain near protosolar
owing to their long diffusion times. {These results identify material originating near the CO$_2$ iceline as the most likely source of Saturn's heavy-element inventory. For the conditions adopted in our model, our results imply that Saturn accreted this material early ($\lesssim0.3$ Myr), in a weakly turbulent disk ($\alpha\lesssim10^{-3}$) with a CO$_2$-rich volatile inventory.}
\end{abstract}



\section{Introduction}\label{sec:intro}

Giant planets in the solar system exhibit a consistent increase in metallicity with heliocentric distance, as evidenced by the inferred bulk heavy-element content from interior models and the measured atmospheric volatile enrichments \citep{Guillot2005,Helled2011,Nettelmann2013,Mousis2022_atmospheres}. Saturn occupies an intermediate position in terms of both semi-major axis and metallicity \citep{Vazan2016,Militzer2019,Movshovitz2020,Ni2020,Mankovich2021}, suggesting that its formation conditions were similar to Jupiter's. Comparisons with Uranus and Neptune are less informative in this context because the ice giants likely followed distinct—and probably later—formation pathways than the gas giants \citep{Dod10,Helled20}. Explaining Saturn’s volatile enrichments is a particularly challenging problem because formation models depend on numerous parameters that directly affect a planet’s final composition \citep{Hubickyj2005,Levison2015,Schneider2021b}.  If Jupiter and Saturn formed through similar mechanisms, then their differing metallicities and volatile inventories must reflect differences in the physical and chemical environments of the disk regions in which they formed.

{The canonical pathway for Saturn's formation is core accretion, in which a solid core grows through the accretion of planetesimals, pebbles, or a combination of both, before acquiring a massive H--He envelope \citep{Pollack1996,Hubickyj2005,Lambrechts2014,Valletta2022}. Within this framework, however, Saturn's detailed growth history remains uncertain. Proposed scenarios differ in the dominant mode of solid accretion, the timing of envelope acquisition, the formation location, and the extent of orbital migration \citep{Levison2015,Schneider2021a,Turrini21,Ronnet20}. Saturn may have formed approximately in situ from locally enriched material, migrated through the disk while accreting solids from a broad radial range, or incorporated volatile-rich planetesimals scattered from other formation regions. The principal alternative to core accretion is gravitational instability, in which a massive disk fragment collapses directly to form a gas giant \citep{Boss1997,Durisen2007}. Such a planet would initially inherit a composition close to that of the local disk gas, and its supersolar metallicity would then require subsequent enrichment by solids, sedimentation, or disk fragmentation in a metal-rich region.}

{The central question is therefore not whether plausible formation mechanisms exist, but which combination of formation location, solid delivery, migration, and disk evolution gave rise to Saturn's present-day heavy-element inventory and atmospheric composition. In this work, we address one component of this broader problem by investigating whether volatile transport near the CO$_2$ or CO icelines can produce material consistent with Saturn's inferred composition.}

In the protosolar nebula (PSN), the volatile budget is primarily controlled by oxygen and carbon, the third and fourth most abundant elements after hydrogen (H) and helium (He). These elements are primarily carried by water (\ce{H2O}), carbon monoxide (CO), and carbon dioxide (\ce{CO2}), which dominate the volatile distribution and the C/O ratio \citep{Woitke2009,Madhusudhan2011,Oberg2011}. {The \ce{H2O} iceline has been widely investigated as a possible location for Jupiter's formation
\citep{Hayashi1981,Drazkowska2017,Schneider2021a,Aguichine2022}, because volatile transport and recondensation can modify the local abundance and composition of solids. However, an iceline is not necessarily a unique or intrinsically preferred site for the formation of a planetary core: core growth may occur wherever the local solid reservoir and dynamical conditions are sufficiently favorable. Moreover, Jupiter's atmospheric composition can also be reproduced at other disk locations if significant orbital migration occurred
\citep{Mousis2019,Oberg2019,Ohno2021}.} 

Similarly, forming Uranus and Neptune near the CO iceline naturally explains their measured D/H ratios \citep{Ali-Dib2014-UN} and elemental abundances \citep{Mousis2020_rspta}. Planetesimals formed in this region and not accreted by the ice giants may share compositional similarities with comet C/2016 R2 \citep{Mousis2021psj} and may also account for the NC/CC chondrite dichotomy \citep{Kruijer2017}. At smaller heliocentric distances, rocklines have also been proposed as the birthplaces of terrestrial planetesimals \citep{Aguichine2020,Izidoro2022}. Because methane is too scarce in the PSN to significantly affect disk composition at its iceline \citep{Garrod2011,Boogert2015,Rubin2019}, the \ce{CO2} iceline remains one of the most plausible formation sites for Saturn (see Fig. \ref{fig:fiducial-icelines}).

{In this work, we investigate whether material processed in the vicinity of the CO or \ce{CO2} iceline could reproduce Saturn's inferred bulk and atmospheric composition. We do not attempt to demonstrate that Saturn's initial core formed at an iceline, nor that an iceline constitutes a uniquely favorable location for the formation of a dominant planetary embryo. Rather, we use the phrase ``formation near an iceline'' to denote the accretion of a substantial fraction of Saturn's heavy-element inventory from the compositional environment generated by volatile transport near that iceline.}

{We therefore perform a quantitative analysis using the 1D disk evolution model of} \cite{Aguichine2020,Aguichine2022}, which tracks the evolution of the \ce{H2}–He gas and the transport of up to 13 species, including 12 volatile carriers of C, N, O, P, S, Ar, Kr, and Xe, and silicate grains. Section~\ref{sec:methods} summarizes the PSN model and highlights the differences relative to \cite{Aguichine2022}. Section~\ref{sec:observables} reviews the compositional constraints on Saturn derived from atmospheric and interior models. In Section~\ref{sec:results}, we use our model results to constrain Saturn’s formation time and location based on its total heavy-element content and to derive implications for its volatile abundances, with particular emphasis on noble gases. {Sections \ref{sec:discussion} and \ref{sec:ccls} are dedicated to Discussion and Conclusion, respectively.}

\section{Methods}\label{sec:methods}

{We investigate Saturn's formation using a one-dimensional model of the evolving PSN that computes the coupled transport of gas, solids, and volatile species. Model predictions are compared with Saturn's atmospheric abundances and inferred envelope metallicity to constrain the formation location and epoch of Saturn's heavy element inventory. The definitions of atmospheric elemental enrichments, planetary metallicity, and their relationship are summarized in Appendix~\ref{app:metallicity}.}

\subsection{PSN model}

{We use the one-dimensional disk evolution model developed by \cite{Aguichine2020,Aguichine2022} to compute the time- and radius-dependent composition of the PSN. The model follows the evolution of the H$_2$--He gas together with the transport, sublimation, recondensation, and radial drift of volatile species and refractory solids. Full details of the numerical implementation are given in \cite{Aguichine2020,Mousis2020_rspta,Aguichine2022}; below we summarize only the aspects relevant to the present study.}

The disk consists primarily of an axisymmetric \ce{H2}–He gas in near-Keplerian rotation and vertical hydrostatic equilibrium, with scale height $H_{\mathrm{g}}$. We consider the disk after infall has ended. The gas surface density $\Sigma_{\mathrm{g}}$ evolves according to \cite{Lynden-Bell1974}:

\begin{equation}
\frac{\partial \Sigma_{\mathrm{g}}}{\partial t} = \frac{3}{r} \frac{\partial}{\partial r} \left[ r^{1/2} \frac{\partial}{\partial r} \left( r^{1/2} \Sigma_{\mathrm{g}} \nu \right) \right],
\label{eq:psn}
\end{equation}

\noindent where $t$ is time and $r$ is the heliocentric distance. The initial condition is given by the self-similar solution of \cite{Lynden-Bell1974}. The disk viscosity is $\nu=\alpha H_{\mathrm{g}}^{2}\Omega_{\mathrm{K}}$, following the $\alpha$-disk formalism \citep{Shakura1973}, where the dimensionless parameter $\alpha$ quantifies turbulent angular-momentum and mass transport. Larger $\alpha$ values lead to faster disk evolution and smoother radial structures. The disk midplane temperature $T$ is obtained from an energy balance including viscous heating, stellar irradiation, and an ambient background temperature $T_{\mathrm{amb}}=10$ K \citep{Hueso2005}. The disk is assumed to be vertically isothermal.

The disk contains 13 trace species: \ce{H2O}, \ce{CO}, \ce{CO2}, \ce{CH3OH}, \ce{CH4}, \ce{N2}, \ce{NH3}, \ce{H2S}, \ce{PH3}, \ce{Ar}, \ce{Kr}, \ce{Xe}, and refractory dust, with abundances listed in Table~\ref{tab:compo-disk}. Each {\it ith} species can exist in either solid or vapor form, with surface densities $\Sigma_{\mathrm{sol},i}$ and $\Sigma_{\mathrm{vap},i}$ evolving according to an advection–diffusion equation \citep{Birnstiel2012,Desch2017}:

\begin{equation}
\frac{\partial \Sigma_{p,i}}{\partial t} + \frac{1}{r} \frac{\partial}{\partial r} \left[ r \left( \Sigma_{p,i} v_{p,i} - D_{p,i} \Sigma_\mathrm{g} \frac{\partial}{\partial r} \left( \frac{\Sigma_{p,i}}{\Sigma_\mathrm{g}} \right) \right) \right] - \dot{Q}_{p,i} = 0,
\label{eq:advdiff}
\end{equation}

\noindent where $p$ denotes the phase (solid or vapor). In the vapor phase, species are advected with the gas ($v_{vap,i}=v_{\mathrm{gas}}$) and diffuse with $D_{vap,i}=\nu$. For solids, the velocity $v_{sol,i}$ includes gas drag and radial drift \citep{Weidenschilling1977,Takeuchi2002,Birnstiel2012}, while turbulent diffusion arises from gas–grain coupling and is treated analogously \citep{Youdin2007,Carballido2011}. Grain Stokes numbers are computed using the two-population model of \cite{Birnstiel2012}, with particle sizes determined in the Epstein or Stokes regimes following \cite{Aguichine2020}. Phase changes are handled through a source–sink term $\dot{Q}_{p,i}$: condensation occurs when the partial pressure exceeds equilibrium vapor pressure, while sublimation follows the Hertz–Knudsen equation \citep{Drazkowska2017}.

\begin{figure}[!h]
\centering
\includegraphics[width=\linewidth]{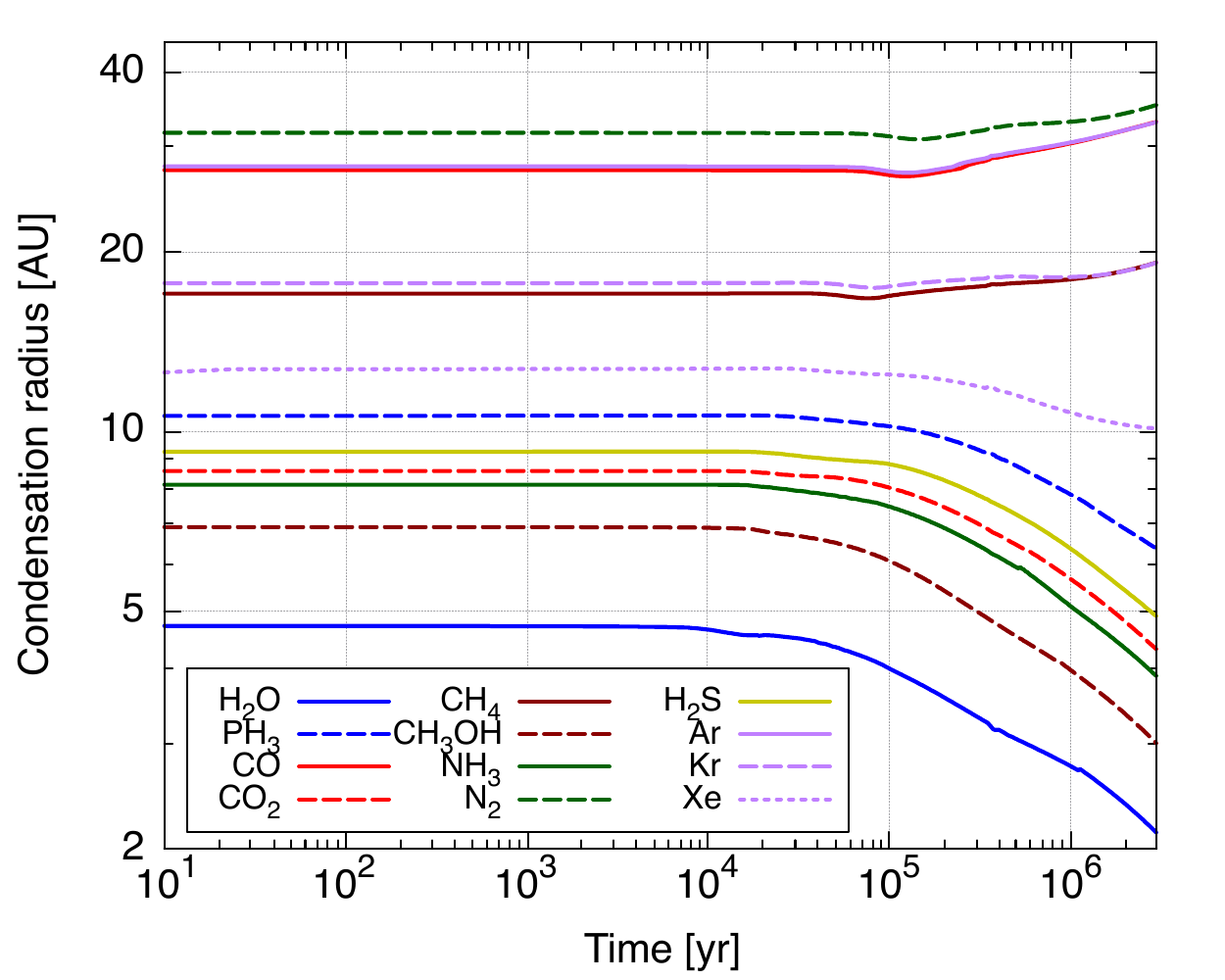}
\caption{{Time evolution} of iceline positions in the PSN for the fiducial case $\alpha=10^{-3}$ and CO:\ce{CO2} = 10:30. Colors and line styles distinguish the 12 volatile species. The relative ordering of icelines is insensitive to the value of $\alpha$.}
\label{fig:fiducial-icelines}
\end{figure}

Unlike models that define icelines by fixed sublimation temperatures, we define the iceline of species $i$ by the distance at which $\Sigma_{\mathrm{sol},i}=\Sigma_{\mathrm{vap},i}$, separating ice- and vapor-dominated regions. Iceline locations therefore depend on temperature, gas pressure, and species abundance. Decreasing disk pressure shifts icelines inward, while lower volatile abundances move them outward. These effects are illustrated in Fig.~\ref{fig:fiducial-icelines}, which shows the time evolution of the 12 volatile icelines for our fiducial case ($\alpha=10^{-3}$, CO:\ce{CO2} = 10:30). {Unlike models that define icelines by fixed sublimation temperatures, our icelines depend on both the disk temperature and gas pressure through the equilibrium between the solid and vapor phases. Consequently, although most icelines migrate inward as the disk cools, the continuing decrease in gas pressure at late times can slightly shift the icelines of the most volatile species outward by lowering their equilibrium condensation temperatures.}

\subsection{Disk parameters and composition}

The initial disk mass is set to $0.1 \msun$ \citep{Hueso2005}, with an initial accretion rate onto the proto-Sun of $10^{-7.6} \msun \mathrm{Myr}^{-1}$ \citep{Ha98}, and an outer radius of 300 AU. We adopt $\alpha$ values between $5\times10^{-4}$ and $5\times10^{-3}$, consistent with disk observations \citep{Ha98,Hueso2005,Desch2017} and with \cite{Aguichine2022}, who showed that this range reproduces Jupiter’s atmospheric composition from PSN material.

At $t=0$, the disk is uniformly populated with 12 volatile species—\ce{H2O}, CO, \ce{CO2}, \ce{CH3OH}, \ce{CH4}, \ce{N2}, \ce{NH3}, \ce{PH3}, \ce{H2S}, Ar, Kr, and Xe—and one refractory dust component. The volatile inventory is based on the dominant ices observed in ISM grains, assumed to approximate the pristine material of the PSN \citep{Oberg2011,Pontoppidan2014,Boogert2015}. All elemental abundances are set to protosolar values \citep{Lodders2009}.

Carbon is partitioned between refractory and volatile reservoirs We assume 50\% of the total carbon is locked in refractory grains, consistent with ISM constraints and Solar System measurements \citep{Oberg2019,Schneider2021a}. The remaining volatile carbon is distributed among CO, \ce{CO2}, \ce{CH3OH}, and \ce{CH4} with ratios CO:\ce{CO2}:\ce{CH3OH}:\ce{CH4} = 10:30:1.67:1. This distribution reflects the dominant carbon-bearing species in ISM ices and is further constrained by ROSINA measurements of comet 67P/Churyumov–Gerasimenko, which indicate substantial CO and \ce{CO2} abundances and minor contributions from reduced carbon species \citep{LeRoy2015,Lauter2019}. Similarly, 20\% of the oxygen is assumed to reside in refractory grains, with the remaining fraction forming oxygen-bearing volatiles. Under this parametrization, \ce{H2O} accounts for 40–60\% of the total oxygen, consistent with the composition of interstellar medium (ISM) ices \citep{Whittet2010,Oberg2011,Boogert2015}. All phosphorus is assumed to reside in \ce{PH3}, while sulfur is partitioned equally between \ce{H2S} and refractory solids \citep{Pasek2005}. Nitrogen is partitioned between \ce{N2} and \ce{NH3} with a ratio of 9:1, consistent with Spitzer observations of molecular cloud cores and protostars \citep{Oberg2011,Pontoppidan2014}. Refractory dust, which neither sublimates nor recondenses, is included as the 13th species. Its abundance satisfies $Z_{\mathrm{tot}} = Z_{\mathrm{ice}} + Z_{\mathrm{dust}} = 0.0155$ \citep{Lodders2009}. Beyond all icelines, the initial solid-to-gas ratio equals $Z_{\mathrm{tot}}$, while interior to the \ce{H2O} iceline it equals $Z_{\mathrm{dust}}$. In this framework, ices account for $\sim$50\% of the pebble mass.

{The adopted volatile partitioning should be regarded as a physically motivated initial condition rather than a unique chemical solution. Astrochemical and thermochemical network models predict that the abundances of CO, CO$_2$,
CH$_4$, NH$_3$, N$_2$, and other species evolve with time as functions of temperature, density, irradiation, ionization, and grain-surface chemistry
(e.g.,  \citealt{Eistrup2016,Eistrup2018,vantHoff2020,Bosman2021}).
Consequently, the volatile inventory of the PSN likely varied both spatially and temporally. Modeling this evolution self-consistently is beyond the scope
of the present study. Instead, we adopt compositions motivated by interstellar ice observations, cometary measurements, and previous PSN studies (e.g., \citealt{Mousis2009,LeRoy2015,Altwegg2019}), and explore a range of CO:CO$_2$ ratios spanning much of the observationally inferred diversity. Temporal evolution of the volatile partitioning would modify the locations and
amplitudes of the enrichment peaks associated with individual icelines. The sensitivity of our conclusions to the adopted CO:CO$_2$ ratio is examined in Sec.~\ref{sec:results}.}

Our fiducial model adopts $\alpha=10^{-3}$, corresponding to moderate turbulence, and CO:\ce{CO2} = 10:30, consistent with Solar System measurements. Because the abundance of \ce{CO2} varies widely among different astrophysical sources, its initial abundance in the PSN is uncertain and may have differed from cometary values. We therefore explore alternative compositions, including CO:\ce{CO2} = 30:10 and the extreme case CO:\ce{CO2} = 40:0 \citep{Boogert2015,Oberg2023}. Higher CO/\ce{CO2} ratios increase the oxygen available for \ce{H2O} formation. {All adopted abundances and the explored compositional ranges are summarized in Appendix~\ref{app:disk}.}


\section{Composition of Saturn} \label{sec:observables}

We adopt the elemental abundances of C, N, and P from \cite{Fletcher2009-ch4}, \cite{Fletcher2011}, and \cite{Fletcher2009-ph3}, respectively. These measurements yield enrichments of C/H $=9.61\pm0.59$, N/H $=2.05\pm1.44$, and P/H $=11.2\pm1.3$ relative to protosolar values (Table~\ref{tab:compo-giants}). The nitrogen abundance is inferred from models that include \ce{NH4HS} cloud formation, which lower the inferred atmospheric N relative to the bulk value.

\begin{figure}
\centering
\includegraphics[width=\linewidth]{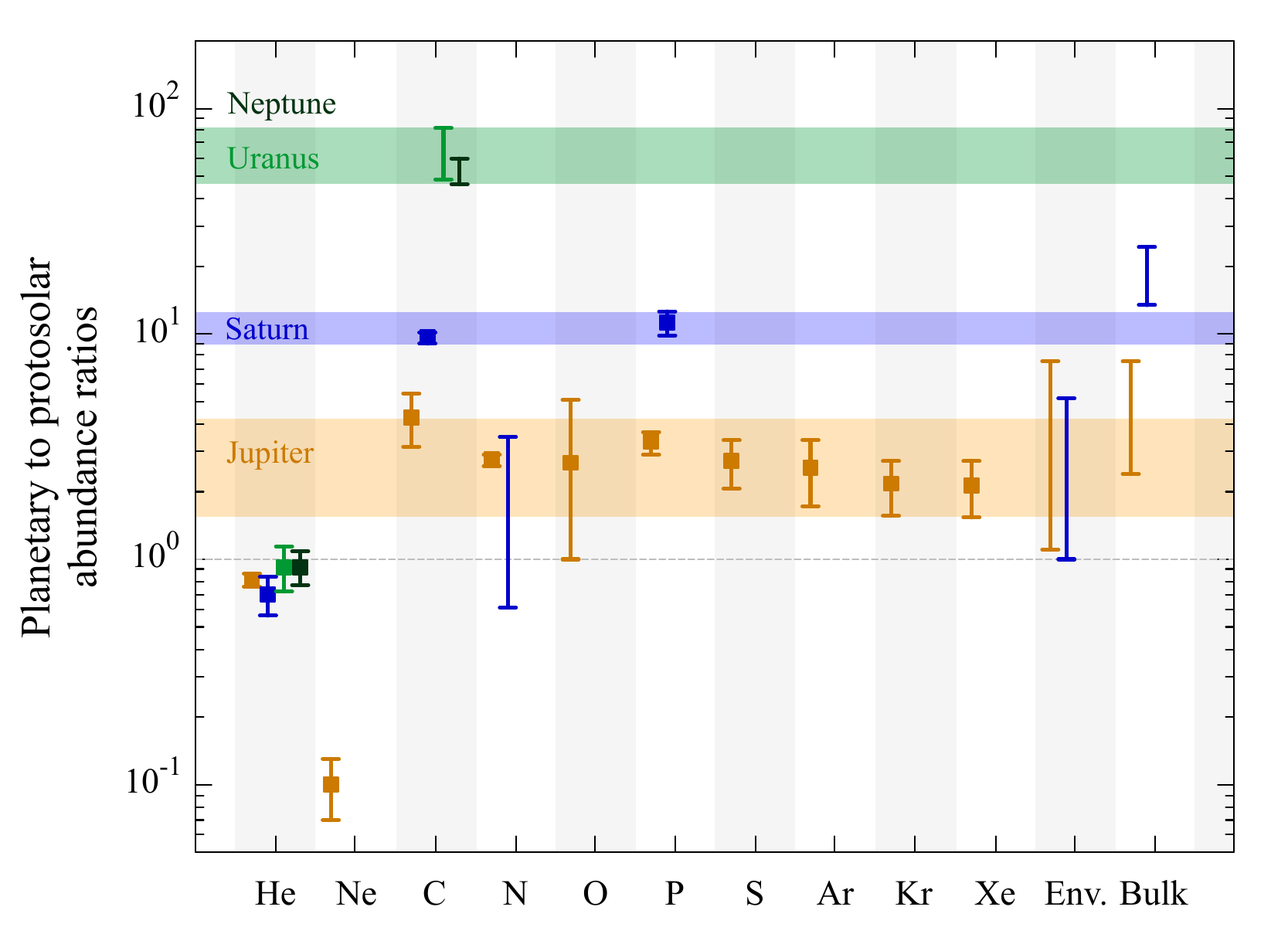}
\caption{Composition of the Solar System’s giant planets. Shaded regions indicate the range of reliably measured atmospheric volatile abundances. Orange denotes Jupiter (Xe to C), blue denotes Saturn (C and P), and green denotes Uranus and Neptune (range of C abundances).}
\label{fig:compo-giants}
\end{figure}

Interior models based on Cassini gravity data infer a total heavy-element mass for Saturn of $M_Z = 16.5$--$26 \mearth$ and core masses of $M_{\mathrm{core}} = 12$–$24 \mearth$ \citep{Militzer2019,Ni2020,Mankovich2021}, corresponding to a bulk metallicity $Z_{\mathrm{bulk}}^\ast \simeq 13.5$--$24.3$. The inferred envelope metallicity, $Z_{\mathrm{env}}^\ast$, spans from near-protosolar up to 4.2, obtained when $M_{\mathrm{core}} = 12 \mearth$ and $M_Z = 16.5 \mearth$. This range ($Z_{\mathrm{env}}^\ast \sim 1$–4.2) is within the envelope metallicities of 1--5.4 inferred by \cite{Nettelmann2021}. Using layer-agnostic interior models, \cite{Movshovitz2020} inferred a higher heavy-element mass of $M_Z = 24$--$32 \mearth$. Although this approach reproduces the density profile without assuming discrete layers, it cannot disentangle heavy-element enrichment from He/H variations, potentially overestimating $Z$. We therefore exclude these results. The enrichments and metallicities of Saturn, and of the other three giants for comparison, are summarized in Table~\ref{tab:compo-giants}. By construction, the envelope metallicity is always lower than the bulk metallicity, as shown in Fig.~\ref{fig:compo-giants}. In the limiting case of a zero-mass core, the envelope and bulk metallicities are identical.

\begin{table*}[]
\centering
\caption{Composition of the Solar System’s giant planets, with atmospheric abundances normalized to protosolar values \citep{Lodders2009}.}
\label{tab:compo-giants}
\begin{threeparttable}
\begin{tabular}{lcccc}
\hline
     & Jupiter                & Saturn        & Uranus        & Neptune       \\ \hline
He/H & $0.81\pm0.05$\tnote{a}          & $0.70\pm0.14$\tnote{b} & $0.93\pm0.21$\tnote{b} & $0.93\pm0.16$\tnote{c} \\
Ne/H & $0.10\pm0.03$\tnote{d}          & -             & -             & -             \\
C/H  & $4.27\pm1.13$\tnote{d}          & $9.61\pm0.59$\tnote{e} & 48--82\tnote{f}        & 46--60\tnote{g}        \\
N/H  & $2.76^{+0.17}_{-0.16}$\tnote{h} & $2.05\pm1.44$\tnote{i} & -             & -             \\
O/H  & $2.70^{+2.40}_{-1.70}$\tnote{h} & -             & -             & -             \\
P/H  & $3.30\pm0.37$\tnote{j}          & $11.2\pm1.3$\tnote{j}  & -             & -             \\
S/H  & $2.73\pm0.65$\tnote{k}          & -             & -             & -             \\
Ar/H & $2.55\pm0.83$\tnote{d}          & -             & -             & -             \\
Kr/H & $2.16\pm0.59$\tnote{d}          & -             & -             & -             \\
Xe/H & $2.12\pm0.59$\tnote{d}          & -             & -             & -             \\
Env. & 1.10--7.50\tnote{l}             & 1--4.2\tnote{m}        & -      & -   \\
Bulk & 2.40--7.50\tnote{l}             & 13.5--24.3\tnote{m}    & -     & -      \\ \hline
\end{tabular}
  \begin{tablenotes}
    \item[a] Galileo \textit{in situ} probe \cite{vonZahn1998,Niemann1998}
    \item[b] Voyager \cite{Conrath1987,Conrath2000,Atreya2018}
    \item[c] Infrared Space Observatory (ISO) \cite{Burgdorf2003}
    \item[d] Galileo \textit{in situ} probe \cite{Mahaffy2000,Wong2004}
    \item[e] Cassini \cite{Fletcher2009-ch4}
    \item[f] HST, Keck, and IRTF \cite{Sromovsky2019}
    \item[g] ALMA \cite{Tollefson2019}
    \item[h] Juno \cite{Li2020}
    \item[i] Cassini \cite{Fletcher2011}; this value likely does not reflect the bulk N abundance
    \item[j] Cassini \cite{Fletcher2009-ph3}
    \item[k] Galileo \textit{in situ} probe \cite{Wong2004}
    \item[l] \cite{Nettelmann2021,Howard2023}
    \item[m] \cite{Militzer2019,Ni2020,Mankovich2021}
  \end{tablenotes}
\end{threeparttable}
\end{table*}

\section{Results}\label{sec:results}

\subsection{PSN enrichment profiles}

\begin{figure}[!h]
\centering
\includegraphics[width=0.4\textwidth]{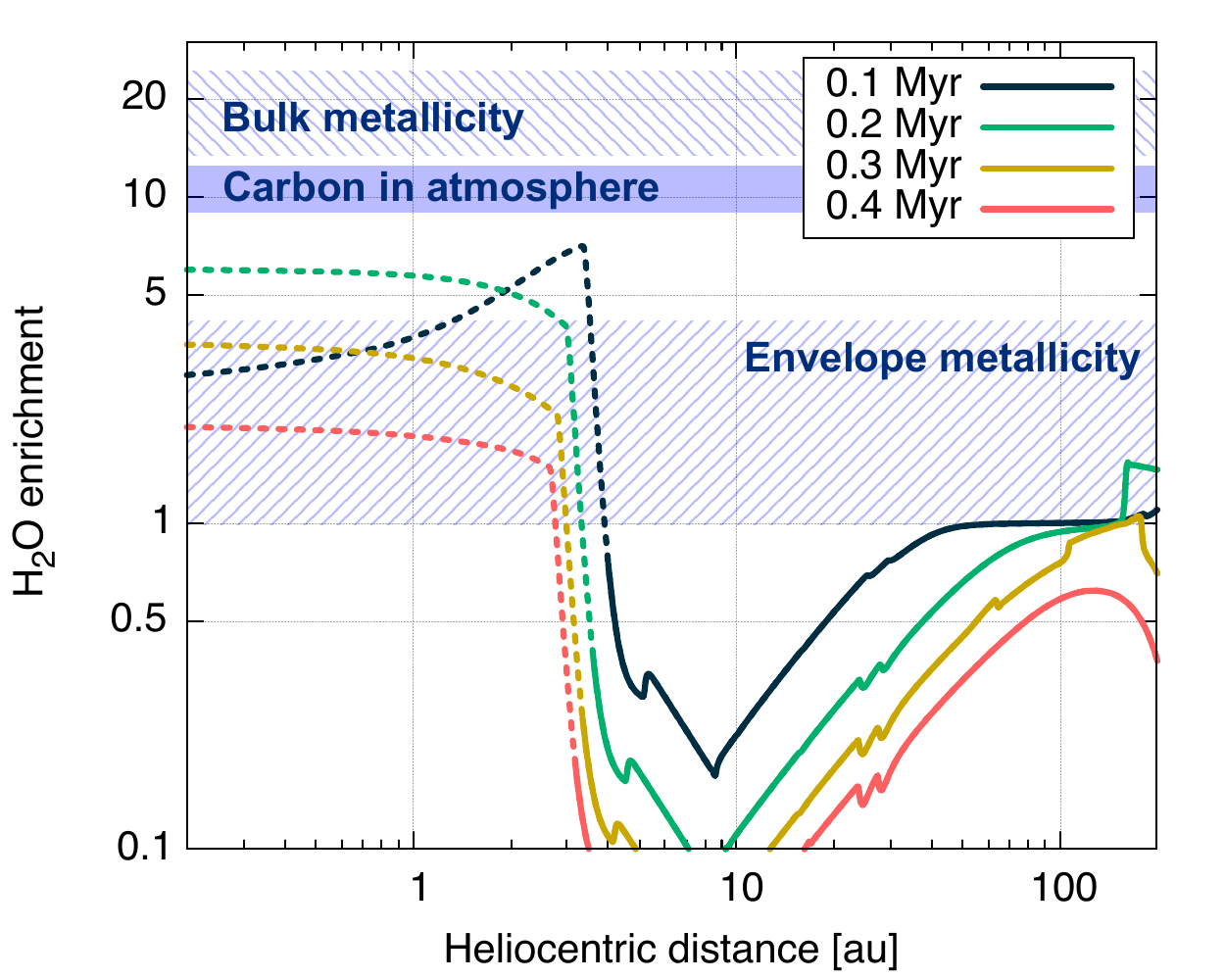}\\[-1mm]
\includegraphics[width=0.4\textwidth]{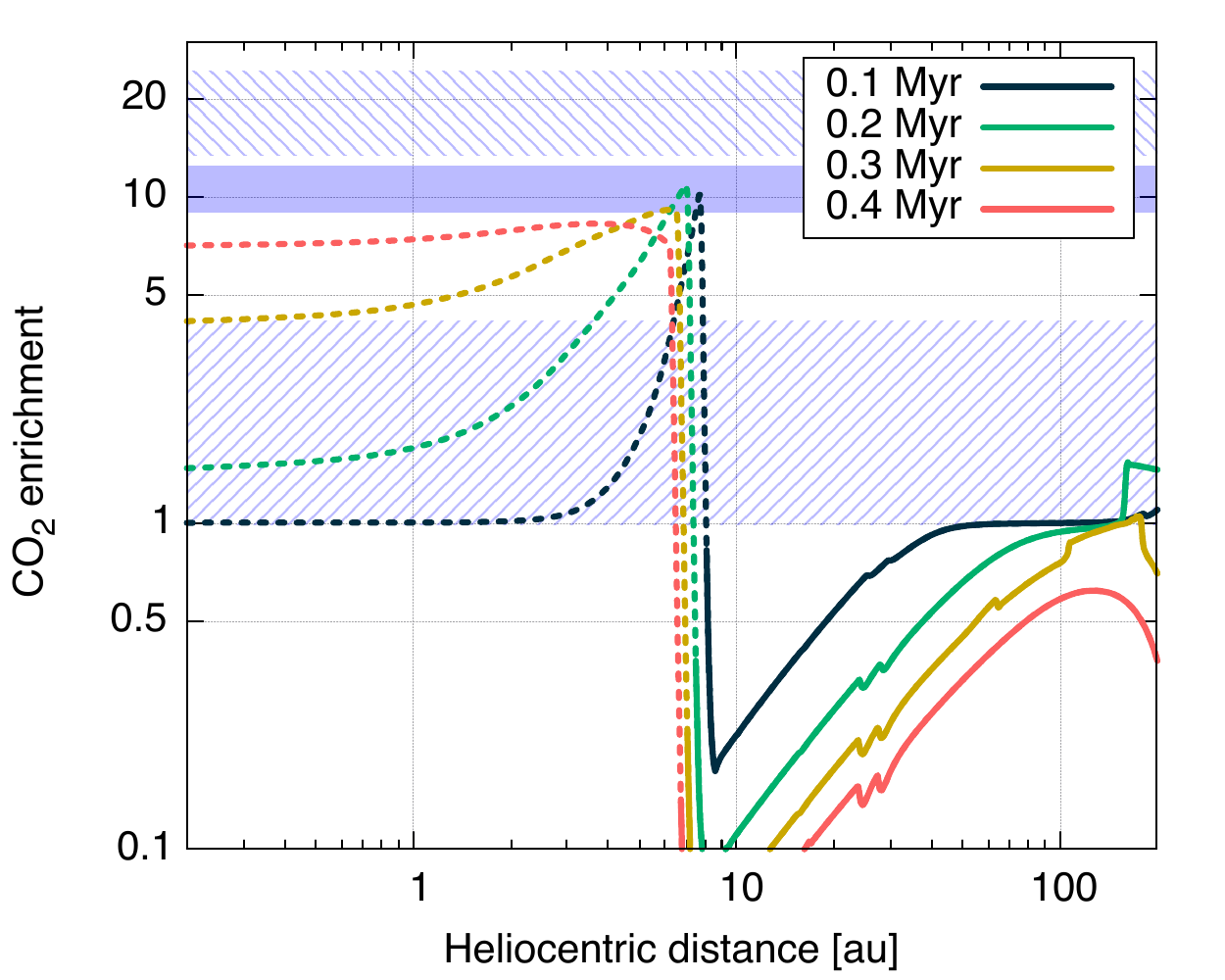}\\[-1mm]
\includegraphics[width=0.4\textwidth]{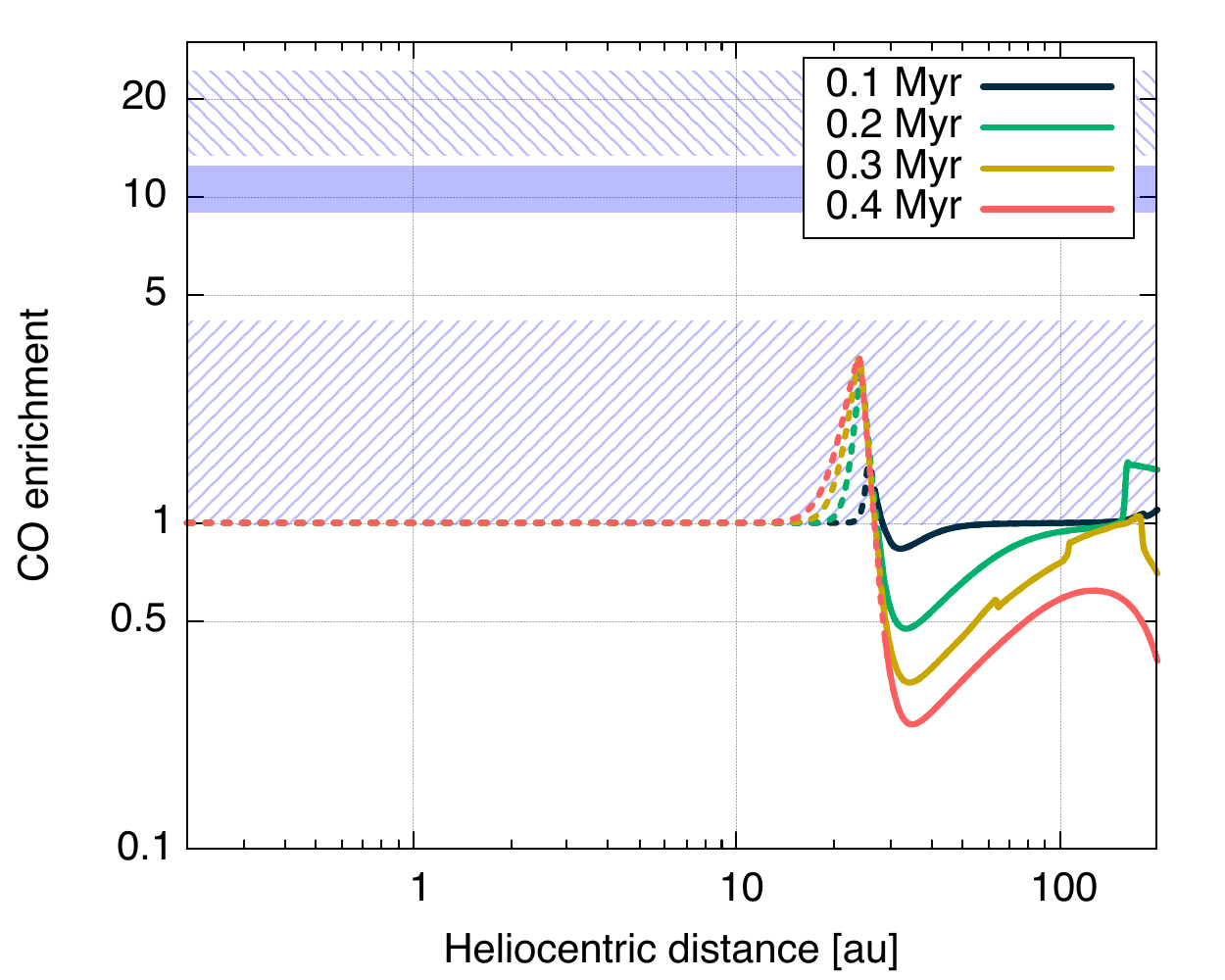}
\caption{Volatile enrichment profiles versus heliocentric distance at multiple stages of disk evolution for the fiducial case ($\alpha = 10^{-3}$ and CO:\ce{CO2} = 10:30). From top to bottom: \ce{H2O}, \ce{CO2}, and CO, ordered according to the locations of their icelines (Fig.~\ref{fig:fiducial-icelines}). Solid lines indicate regions where the species is predominantly solid ($\Sigma_{\mathrm{solid},i} > \Sigma_{\mathrm{vapor},i}$), while dashed lines indicate vapor-dominated regions ($\Sigma_{\mathrm{solid},i} < \Sigma_{\mathrm{vapor},i}$). In each panel, the lower shaded, central filled, and upper shaded bands denote Saturn’s envelope metallicity, atmospheric enrichment considering C and P values only, and bulk metallicity, respectively. Agreement with Saturn’s atmospheric enrichment occurs where dashed curves intersect the central band, here only near the \ce{CO2} iceline.}
\label{fig:fiducial-profiles}
\end{figure}

Figure~\ref{fig:fiducial-profiles} shows the enrichment profiles of \ce{H2O}, \ce{CO2}, and \ce{CO} as functions of heliocentric distance at several epochs of disk evolution. The inward drift of icy grains progressively depletes solids in the outer PSN until each species reaches its iceline and sublimates. The resulting vapors diffuse both inward and outward, but more slowly than the inward drift of solids. As a consequence, vapors accumulate at the inner edge of each iceline, while outward diffusion allows partial recondensation, further reducing net inward volatile transport.

These combined processes generate enrichment peaks at each iceline—the so-called cold-finger effect \citep{Lichtenberg2021}. In our model, vapor diffusivity is assumed equal to the gas viscosity, leading to efficient mixing and the formation of an inner disk region with nearly constant vapor-to-gas ratio, referred to as an enrichment plateau. 

{This assumption corresponds to adopting a turbulent Schmidt number $Sc \equiv \nu/D_{\mathrm{vap}} = 1$, where $D_{\mathrm{vap}}$ is the vapor mass diffusivity and $\nu$ denotes the effective turbulent viscosity \citep{Armitage2020}. The value of $Sc$ is uncertain and may differ from unity. At fixed viscosity, $Sc>1$ implies weaker vapor diffusion, which would confine vapor more closely to its iceline and delay the establishment of enrichment plateaus at smaller heliocentric distances. Conversely, $Sc<1$ would produce faster and more extended vapor redistribution. Variations in $Sc$ would also affect the fraction of vapor that diffuses outward across an iceline and recondenses, thereby modifying the strength of the cold-finger effect. Because stronger diffusion simultaneously enhances outward recondensation and dilutes local vapor enhancements, the net effect on the peak amplitude is not necessarily monotonic. \cite{Yap2026} demonstrated a strong dependence of volatile partitioning on $Sc$ in a decreting circumplanetary disk. Although the transport geometry in that model differs from the globally
evolving, accreting PSN considered here, their results emphasize that $Sc$ is an important source of uncertainty. Our quantitative enrichment factors and diffusion timescales should therefore be understood as conditional on $Sc=1$; exploring this parameter systematically is left
for future work.}

Because diffusion times increase with distance, more distant icelines produce plateaus later in time; for example, CO vapors require $\sim$1.3 Myr to reach the inner disk. Finally, trace species are dynamically coupled because the presence of volatile ices controls grain growth and particle sizes, which in turn set the radial drift velocities transporting all condensed species.

Figure~\ref{fig:fiducial-profiles} further shows that Saturn’s carbon enrichment is reproduced only near the \ce{CO2} iceline, specifically when \ce{CO2} is in vapor form, where inward drift, vapor diffusion, and partial recondensation yield consistent volatile enhancements.

\subsection{Metallicity of the PSN}

\begin{figure*}[!h]
    \centering
    \includegraphics[width=0.99\textwidth]{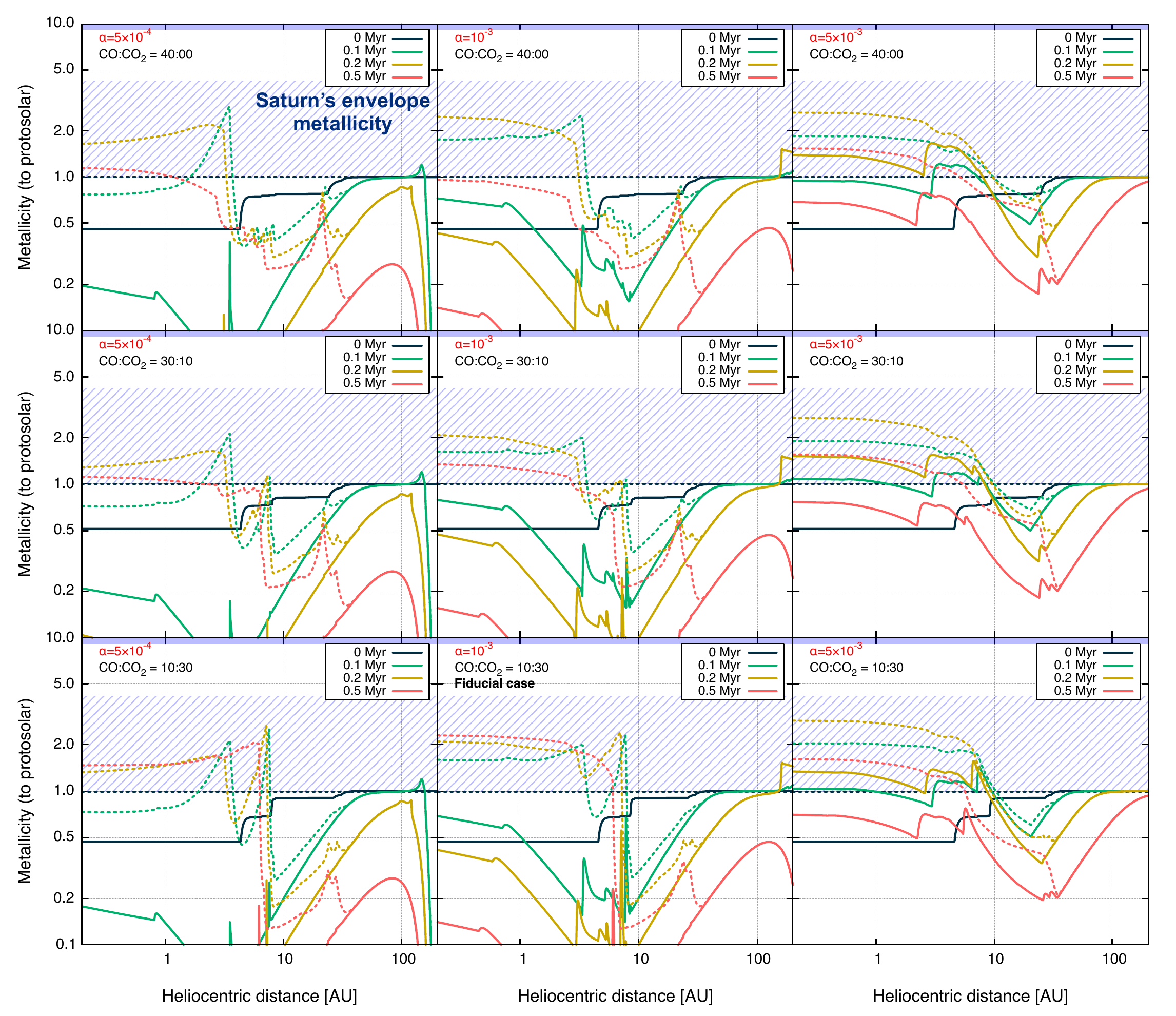}
    \caption{PSN metallicity profiles as a function of heliocentric distance at several stages of disk evolution, normalized to the protosolar value. Columns (left to right) show $\alpha$ = $5\times 10^{-4}$, $10^{-3}$, and $5\times 10^{-3}$; rows (top to bottom) correspond to CO:\ce{CO2} ratios of 40:00, 30:10, and 10:30. The fiducial case is the center panel of the bottom row. Solid lines include solids only, while dashed lines include both solids and vapor. Shaded regions are as in Fig.~\ref{fig:fiducial-profiles}. Matching Saturn’s envelope metallicity requires accretion of gas and solids interior to 10 AU.}
    \label{fig:metallicity-all}
\end{figure*}

Figure~\ref{fig:metallicity-all} shows metallicity profiles normalized to protosolar values at different epochs of PSN evolution, for $\alpha = 5\times10^{-4}$, $10^{-3}$, and $5\times10^{-3}$, and CO:\ce{CO2} ratios of 40:00, 30:10, and 10:30. The profiles represent the summed contributions of the 13 trace species. Solid lines include only solids, while dashed lines include both solids and vapors.

At $t=0$, all species are uniformly mixed, yielding a constant solid+vapor metallicity of unity. In the outer disk, where all volatiles are condensed, the solid metallicity is protosolar. Moving inward, successive sublimation at icelines transfers material to the vapor phase, and interior to the innermost iceline only refractory grains remain, setting the metallicity to $\sim$0.5.

As the disk evolves, each volatile produces a metallicity peak at its iceline via the cold-finger effect, with peak amplitudes scaling with species abundance. Noble-gas signatures remain weak because of their low abundances, whereas strong metallicity peaks appear at the icelines of \ce{H2O} ($\sim$4 AU), \ce{NH3} ($\sim$7 AU), \ce{CO2} ($\sim$8 AU), CO ($\sim$25 AU), and \ce{N2} ($\sim$30 AU). An additional metallicity enhancement at $\sim$0.5 AU, present for $\alpha \leq 10^{-3}$, is unrelated to the cold-finger effect and instead results from a transition in the opacity regime that modifies the pressure gradient and pebble drift. This feature coincides with the location of rocklines \citep{Aguichine2020} and affects only the inner disk, with no impact on giant-planet composition.

Increasing $\alpha$ enhances turbulent mixing, producing broader and weaker metallicity peaks. Lower CO/\ce{CO2} ratios reduce the CO peak at $\sim$25 AU and, by limiting available oxygen, also weaken the \ce{H2O} peak at $\sim$4 AU. The total PSN metallicity (solids + vapors) matches Saturn’s envelope metallicity (1–4.2$\times$ protosolar) only under the following conditions:

\begin{itemize}
    \item  $\alpha \leq 10^{-3}$, CO:\ce{CO2} = 40:00: interior to the \ce{H2O} iceline at 0.1--0.4 Myr;
    \item $\alpha \leq 10^{-3}$, CO:\ce{CO2} = 30:10 or 10:30: interior to the \ce{H2O} iceline or near the \ce{CO2} iceline at 0.1–0.5 Myr;
    \item $\alpha = 5\times10^{-3}$: within $\sim$10 AU, slightly beyond the \ce{CO2} iceline.
\end{itemize}

Saturn’s metallicity can always be reproduced interior to the \ce{H2O} iceline. However, this location conflicts with scenarios in which Jupiter formed near the \ce{H2O} iceline \citep{Aguichine2022}. Forming both planets within the same framework therefore favors Saturn’s growth between the \ce{H2O} and \ce{CO2} icelines. Notably, the total disk metallicity is never supersolar at the CO iceline, even when carbon is entirely in the form of CO. This behavior arises because the CO iceline lies beyond the centrifugal radius, defined here as the radius where the gas radial velocity changes sign from inward (negative) to outward (positive). The centrifugal radius is computed self-consistently and evolves from 1.9 AU at $t=0$ to 22 AU at $t=2$ Myr for $\alpha=10^{-3}$; beyond it, the CO abundance gradient is shallower and diffusion timescales are longer. As a result, vapor accumulation interior to the CO iceline is limited, reducing the efficiency of the cold-finger effect. Combined with the fact that the CO iceline lies at $\sim$30 AU (Fig.~\ref{fig:fiducial-icelines}), well beyond Saturn’s likely formation region, these results strongly favor Saturn’s formation near the \ce{CO2} iceline and imply that a minimum \ce{CO2} abundance is required to match its envelope metallicity in a core-accretion scenario where solids and gas are accreted simultaneously.

{Although our calculations do not include time-dependent chemical evolution, the explored range of CO:CO$_2$ ratios encompasses much of the variability inferred from interstellar and cometary observations. Within this parameter space, changes in the initial volatile partitioning primarily modify the amplitudes of the carbon and oxygen enrichment peaks, while leaving the overall evolution of the enrichment profiles unchanged. The preference for Saturn's formation near the CO$_2$ iceline should be robust even accross other volatile compositions than those considered here.}

In contrast, the high bulk metallicities of Uranus and Neptune likely result from enhanced solid accretion relative to gas, via pebble, planetesimal, or hybrid (pebbles + planetesimals) accretion \citep{Lambrechts2012,Lambrechts2014,Valletta2022,Malamud2024,Eriksson2023}, disk dissipation \citep{Ercolano2009,Owen2012-disk-evap,Owen2013-disk-evap}, or accretion of inward-migrating planetary embryos \citep{Izidoro2015}. Consequently, their present-day metallicities should not be viewed as representative of the intrinsic metallicity of the PSN. {Finally, the radial distribution and magnitude of the solid enhancements depend strongly on $\alpha$. In Section~\ref{sec:formation-pt}, we use a simplified pebble-to-planetesimal conversion prescription to estimate the amount and radial distribution of material that could be incorporated into planetesimals. This calculation is intended as a diagnostic of the available solid reservoir and not as a self-consistent model of streaming instability.}

\subsection{Volatile enrichments at the \ce{CO2} iceline}

\begin{figure*}[!h]
\centering
\includegraphics[width=0.99\textwidth]{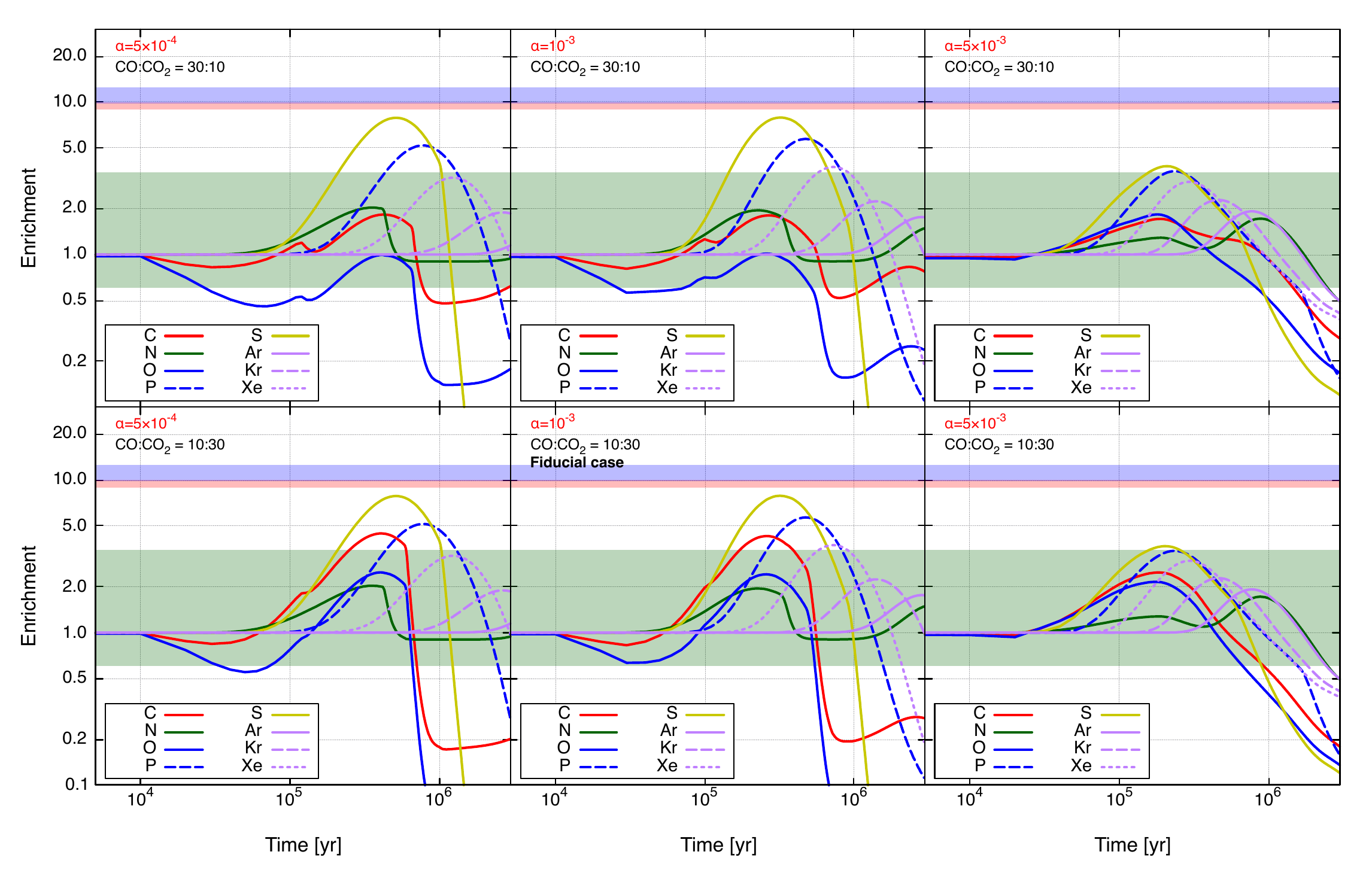}
\caption{Time evolution of volatile enrichments (solids + vapor) at 6 AU. The six panels are a subset of Fig.~\ref{fig:metallicity-all}, excluding CO:\ce{CO2} = 40:00. Rows correspond to CO:\ce{CO2} = 30:10 (top) and 10:30 (bottom), and columns to increasing $\alpha$. The fiducial case is the center panel of the bottom row. Red, blue, and green bars show the C, P, and N enrichments in Saturn’s atmosphere inferred from Cassini measurements \citep{Fletcher2009-ph3,Fletcher2009-ch4,Fletcher2011}.}
\label{fig:enrichment-all}
\end{figure*}

Based on the previous results, we exclude the case CO:\ce{CO2} = 40:00, as it fails to reproduce metallicities comparable to those inferred for Saturn’s interior at the \ce{CO2} iceline (Fig. \ref{fig:metallicity-all}). Figure~\ref{fig:enrichment-all} shows the time evolution of elemental enrichments at 6 AU, just interior to the location of the \ce{CO2} iceline, which remains close to this distance during the first 1 Myr of PSN evolution. 

For elements carried by a single volatile species (P, S, Ar, Kr, and Xe), enrichments rise from protosolar to a maximum between 0.2 and 1.0 Myr as vapors released at icelines diffuse inward to 6 AU. More distant icelines produce later peaks. In contrast, C, N, and O exhibit multiple enrichment maxima reflecting contributions from several volatile carriers. For N, the first peak is associated with \ce{NH3} ($\sim$0.3 Myr) and the second with \ce{N2} ($\sim$1.0 Myr). For C and O, a minor early peak arises from \ce{CH3OH} ($\sim$0.1 Myr), followed by a dominant contribution from \ce{CO2} ($\sim$0.3 Myr), with CO contributing at 1--3 Myr depending on $\alpha$. \ce{CH4} contributes weakly and is not visible at the adopted scale. At times $<$ 0.1 Myr, C and O are subsolar due to depletion of refractory material and \ce{H2O} ice at 6 AU. Because a large fraction of oxygen is locked in \ce{H2O}, O enrichments are generally lower than those of C.

Figure \ref{fig:enrichment-all} also shows that increasing the CO:\ce{CO2} ratio from 10:30 to 30:10 enhances the C and O enrichment peaks associated with \ce{CO2} at $\sim$0.3 Myr, while making the CO-driven features at $\sim$2 Myr less prominent, as expected. Other elemental enrichment profiles remain essentially unchanged. Increasing $\alpha$ shifts enrichment peaks to earlier times and compresses them temporally due to more efficient turbulent diffusion, but also reduces their amplitudes as a result of enhanced mixing. These trends are consistent with the results of \cite{Aguichine2022}.

\begin{figure*}[!h]
\centering
\includegraphics[width=0.95\textwidth]{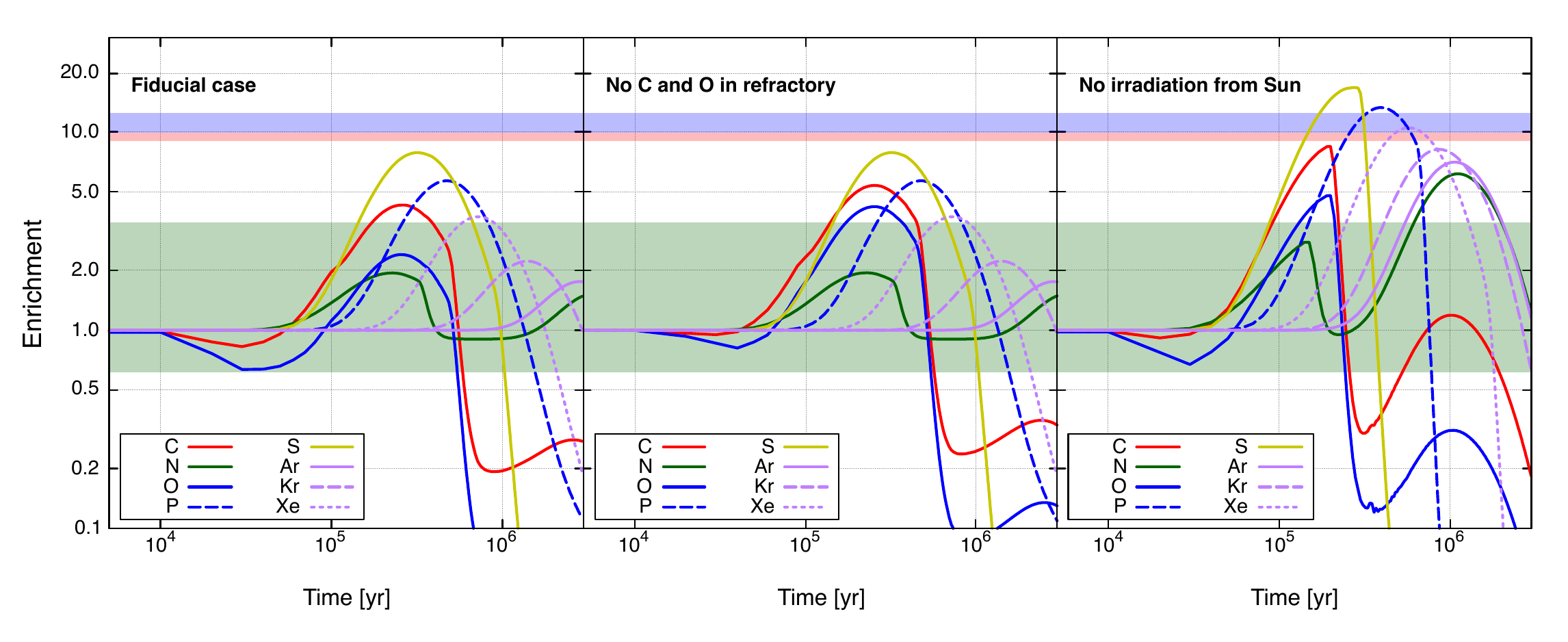}
\caption{Same as Fig.~\ref{fig:enrichment-all}, but for the fiducial case (left) and two derived variants: one assuming no C and O in minerals (center), and one neglecting solar irradiation (right). Only the case without solar irradiation enables our model to simultaneously reproduce the observed C and P enrichments.}
\label{fig:enrichment-fiducial-compare}
\end{figure*}

To test the robustness of our results, we explore two variants of the fiducial model: (i) all carbon and oxygen are assumed to reside in volatile species (i.e., no C and O in minerals), and (ii) solar irradiation is neglected. Neglecting irradiation is motivated by the possible shadowing of the outer disk by its inner regions \citep{Ohno2021}. Results for these cases are shown in Fig.~\ref{fig:enrichment-fiducial-compare}. Because enrichment peaks arise from icelines, allocating more C and O to ices enhances their enrichments; however, a 50\% increase in C locked in ices yields only a 20\% increase in C enrichment (middle panel). In contrast, the results are far more sensitive to the absence of solar irradiation. In this case, viscous heating dominates and is confined to the inner disk, producing a steeper temperature profile that places all icelines within $\sim$13 AU, compared to $\sim$31 AU in the fiducial case (Fig.~\ref{fig:fiducial-icelines}). When icelines are closer to the Sun, a larger reservoir of ice exterior to the iceline can drift inward, strengthening the cold finger effect. Thus, enrichment is governed primarily by the iceline’s position relative to the disk mass distribution rather than by its absolute heliocentric distance.

The strongest carbon enrichment occurs just interior to the \ce{CO2} iceline, where \ce{CO2} vapor accumulates via the cold-finger effect (Fig.~\ref{fig:fiducial-profiles}). However, the carbon abundance shown in Fig.~5 is not reproduced by the model. Reproducing Saturn’s atmospheric carbon abundance (C/H $=9.61\pm 0.59$ times protosolar) requires most carbon to be carried by \ce{CO2}, implying a predominantly volatile carbon inventory. Disk viscosities $\alpha \le 10^{-3}$ are favored, and only models without solar irradiation simultaneously reproduce the observed C and P enrichments at 6 AU and $\sim$0.2 Myr. Nitrogen abundances are always consistent with Saturn’s measurements, largely due to their large uncertainties, suggesting that N alone provides weak constraints on pebble-driven formation scenarios. Including planetesimal accretion could modify this conclusion.

\subsection{Compatibility with planetesimal accretion} \label{sec:formation-pt}

{So far, we have focused on whether the redistribution of pebbles and volatile vapors can generate material whose composition is compatible with Saturn's envelope. Our model does not attempt to describe the formation of Saturn's initial core or to establish that an iceline is a uniquely favorable location for core formation. Core formation may occur wherever the local solid reservoir and dynamical conditions permit sufficiently rapid growth, and determining the subsequent evolution of planetesimals and planetary embryos is beyond the scope of this work.}

{For completeness, we nevertheless estimate the amount and radial distribution of material that could be converted from pebbles into planetesimals. This calculation should be regarded as a diagnostic of the available solid reservoir rather than as a self-consistent model of streaming instability. In particular, the parameter $\alpha$ in our calculations controls viscous evolution and turbulent transport of gas and trace species. A self-consistent treatment of streaming instability would additionally require modeling the dependence of particle settling, midplane dust-to-gas ratios, and instability thresholds on turbulence. We do not attempt such a calculation here. Instead, we adopt the simplified planetesimal-conversion prescription of \cite{Drazkowska2017} to assess whether the disk contains a planetesimal mass broadly compatible with Saturn's heavy-element inventory.}

Following \cite{Drazkowska2017}, we define the planetesimal formation rate as

\begin{equation}
\dot{\Sigma}_{\mathrm{plt}} = \zeta \Sigma_{solid}\left(\mathrm{St} > 10^{-2}\right) \OmegaK, \label{eq:sigma-pt-dot}
\end{equation}

\noindent where $\zeta=10^{-3}$ is an assumed pebble-to-planetesimal conversion efficiency \citep{Drazkowska2016,Simon2016}. {The adopted efficiency $\zeta$ is therefore phenomenological. In particular, we do not calculate whether the local particle concentration satisfies a
turbulence-dependent streaming-instability threshold. The resulting planetesimal surface densities quantify where the adopted conversion prescription processes the pebble reservoir most efficiently, but they should not be interpreted as demonstrating that these locations necessarily produce
a single massive planetary core.}

At each time step $\Delta t$ and radial location, a mass $\dot{\Sigma}_{\mathrm{plt}},\Delta t$ is removed from the pebble reservoir (i.e., the largest particles in the adopted two-population model) represented by $\dot{Q}$ (Eq.~\ref{eq:advdiff}) and converted into planetesimals. Planetesimals inherit the composition of their parent pebbles, ensuring conservation of total mass and of each trace species. This prescription allows us to track the mass and composition of planetesimals as functions of time and heliocentric distance. In this simplified framework, planetesimal sizes and mutual interactions are neglected; their orbits are assumed circular and static, with no dust sweeping or collisional recycling. This approach is designed to address two questions: the timescale of planetesimal formation relative to the PSN lifetime, and the disk regions most favorable to their formation.

\begin{figure*}[!h]
\centering
\includegraphics[width=0.8\textwidth]{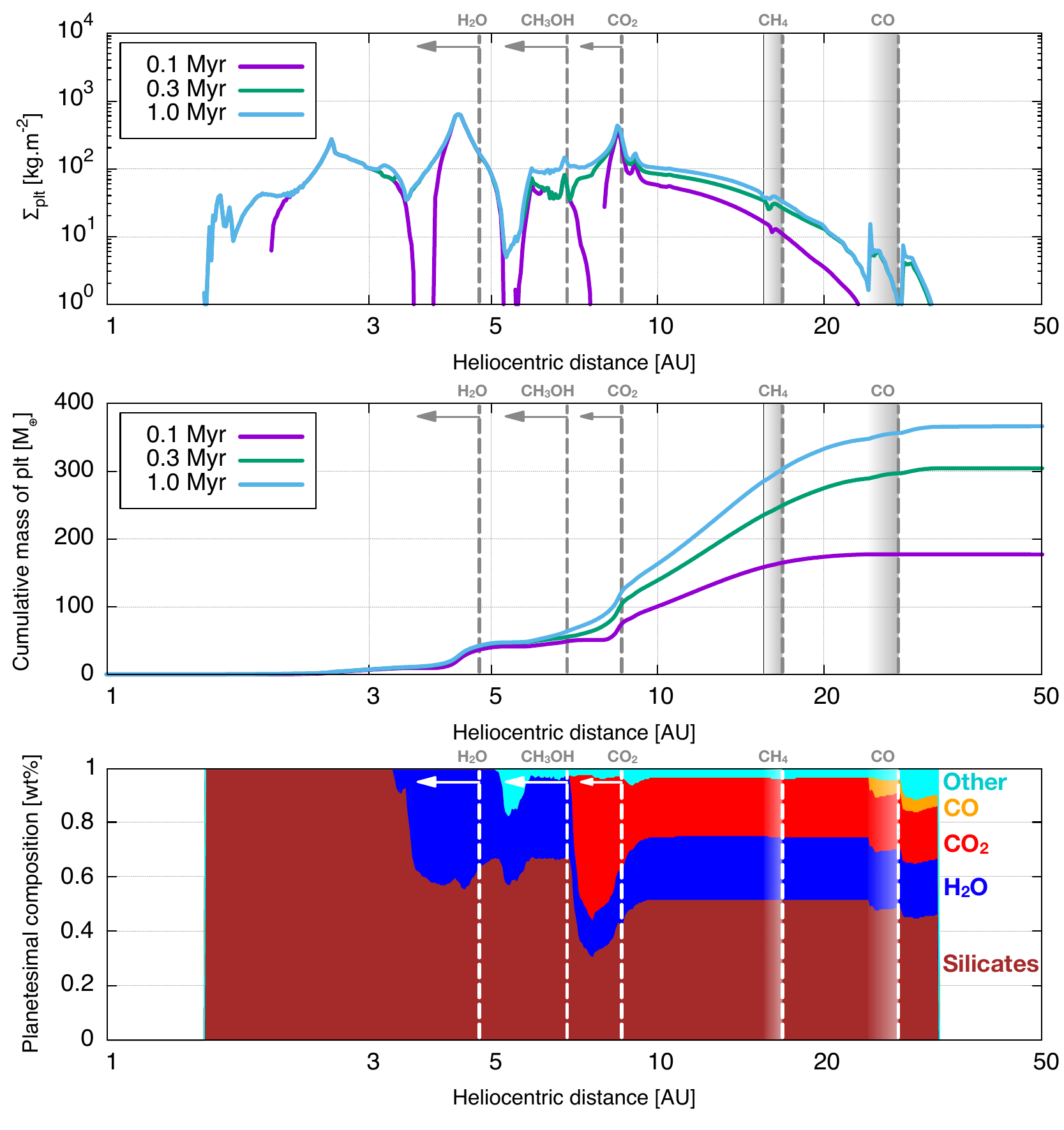}
\caption{Results of the fiducial run including planetesimal formation (Eq.~\ref{eq:sigma-pt-dot}). The top panel shows planetesimal surface density profiles at several stages of disk evolution, with enhanced formation near icelines within 2--30 AU. {The middle panel shows the cumulative mass converted into planetesimals by the adopted prescription as a function of heliocentric distance. Because subsequent planetary accretion and dynamical depletion are not included, these values represent the cumulative formed planetesimal mass and should not be interpreted as the surviving planetesimal inventory.} The bottom panel shows the mean planetesimal composition (mass fraction) at 1 Myr, with colors indicating individual trace species. Vertical dashed lines mark the initial locations of the \ce{H2O}, \ce{CH3OH}, \ce{CO2}, \ce{CH4}, and CO icelines (left to right), arrows indicate their positions at 0.3 Myr, and shaded regions denote zones where vapor and ice coexist.}
\label{fig:planetesimals}
\end{figure*}

We use the same numerical setup as in the fiducial case, with the addition of Eq.~(\ref{eq:sigma-pt-dot}). The resulting planetesimal surface-density profiles are shown in Fig.~\ref{fig:planetesimals}. {The adopted prescription produces enhanced pebble-to-planetesimal conversion in regions of pebble pile-up \citep{Drazkowska2016,Drazkowska2017} and near some icelines \citep{Schoonenberg2018}. These enhancements identify regions where the modeled pebble reservoir is locally concentrated; they do not constitute a self-consistent calculation of the onset of streaming instability.} Formation in local dust traps \citep{Lenz2020} is not captured, as such structures are absent from the modeled gas disk. Figure~\ref{fig:planetesimals} shows the surface density of planetesimals as a function of heliocentric distance and time. Planetesimal formation begins at $\sim$0.01 Myr, once particles reach Stokes numbers $\gtrsim10^{-2}$, and continues throughout the disk lifetime wherever dust is present. {Within the adopted conversion prescription, planetesimals are produced between 2 and 30 AU, with enhanced conversion near some icelines because of locally increased pebble surface densities and Stokes numbers.} Because icelines migrate with time, planetesimal surface-density peaks do not always coincide with their initial locations, notably for \ce{H2O}, \ce{CH3OH}, and \ce{CO2}. For species whose icelines remain nearly stationary, such as \ce{CH4} and CO, planetesimal peaks occur slightly interior to the iceline, reflecting the extended radial region where solids and vapors coexist. {The highest planetesimal surface densities predicted by this prescription are reached near the locations where the final major solid components sublimate.} Interior to the \ce{H2O} iceline, planetesimals form exclusively by pebble pile-up, with the peak at $\sim$2.5 AU arising from a local maximum in grain size caused by the transition from fragmentation-limited to drift-limited growth \citep{Birnstiel2012,Aguichine2020}.

The middle panel shows the cumulative planetesimal mass as a function of heliocentric distance. From an initial trace-species budget of $\sim520 \mearth$, the model forms $\sim175 \mearth$ of planetesimals by 0.1 Myr, $\sim300 \mearth$ by 0.3 Myr, and $\sim360 \mearth$ by 1.0 Myr. While the formation timescale is consistent with \cite{Drazkowska2017}, the total mass is higher because planetesimals form at multiple disk locations rather than only near the \ce{H2O} iceline. At 1 Myr, $\sim5 \mearth$ of planetesimals form within 3 AU from volatile-free dust, $\sim35 \mearth$ between 3 and 5 AU near the \ce{H2O} iceline, sufficient to form Jupiter’s core, and $\sim120 \mearth$ between 5 and 10 AU near the \ce{CH3OH} and \ce{CO2} icelines, exceeding Saturn’s heavy-element budget by a factor of 4--6. The remaining $\sim200 \mearth$ of planetesimals lie beyond 10 AU. {These cumulative masses should not be interpreted as the surviving planetesimal inventory of the Solar System. Rather, they represent the total mass converted from pebbles into planetesimals by the adopted prescription. The present model does not include subsequent planetesimal accretion by growing planets, gravitational scattering, collisional evolution, or dynamical depletion, all of which would substantially reduce the remaining planetesimal population.}

Figure~\ref{fig:planetesimals} also shows the average planetesimal composition at 1 Myr. Planetesimals formed within 3 AU, interior to the \ce{H2O} iceline, are volatile-free. Between 3 and 5 AU, they contain up to $\sim$40\% \ce{H2O} by mass. At larger heliocentric distances, additional species condense, leading to increasingly diverse compositions; beyond 10 AU, the volatile fraction approaches the protosolar value ($\sim$53\% by mass). Supersolar volatile contents occur at the \ce{CO2} iceline due to the cold-finger effect. There, planetesimals consist of $\sim$31\% silicates and 69\% volatiles, dominated by \ce{CO2} (53\%), with smaller contributions from \ce{H2O} (14\%) and other species (2\%). 

To explain the presence of ammonia on Ceres’ surface \citep{deSanctis2015}, recent dynamical models of the early Solar System show that Ceres-sized planetesimals could migrate from the trans-Saturnian disk—where \ce{NH3} condenses—to Ceres’ current orbit, interior to the \ce{NH3} iceline \citep{RibeirodeSousa2022}. Although planetesimal dynamics are beyond the scope of this work, we discuss the compatibility of this scenario with our results below. The \ce{NH3} iceline lies $\sim$1 AU interior to the \ce{CO2} iceline (Fig.~\ref{fig:fiducial-icelines}), implying that if Saturn formed near the \ce{CO2} iceline, planetesimals originating in the trans-Saturnian disk would naturally incorporate \ce{NH3}, as observed for Ceres. However, such planetesimals would also be expected to be ice-rich, with rock-to-ice ratios near unity or lower, inconsistent with Ceres’ bulk composition \citep{McCord2005,Zolotov2020}. Reconciling these properties would therefore require additional processes, such as collisional stripping, to reduce the volatile content of Ceres \citep{Schlichting2015,Izidoro2022}.

\section{Discussion}\label{sec:discussion}

\subsection{Atmospheric constraints on Saturn's formation}

{Carbon provides the strongest compositional diagnostic because its abundance is controlled primarily by the relative contributions of CO$_2$ and CO. Taken together, the atmospheric abundances and inferred envelope metallicity are most naturally reproduced by material originating in the vicinity of the CO$_2$ iceline. Oxygen is generally predicted to be less enriched than carbon, yielding a supersolar atmospheric C/O ratio, approximately 1.2--2 times the protosolar value depending on the formation epoch, similar to that inferred for Jupiter \citep{Li2020,Cavalie2023}.}

{Nitrogen is a weaker discriminator because its predicted enrichment depends primarily on the poorly constrained NH$_3$/N$_2$ partitioning
\citep{Oberg2011,Pontoppidan2014,Fegley2000,Mousis2009-titan}. A larger initial NH$_3$ fraction would produce substantially larger atmospheric nitrogen enrichments than predicted by our fiducial model.}

{Conversely, the noble gases provide one of the strongest observational tests of the proposed formation scenario. Because their icelines lie much farther from the Sun than that of CO$_2$, their enrichments develop on substantially longer diffusion timescales. As a result, Ar and Kr remain close to protosolar abundances during the epoch when Saturn's carbon and phosphorus enrichments are reproduced, while Xe is predicted to become only slightly supersolar. Future \textit{in situ}
measurements by a Saturn entry probe would therefore provide a direct test of the timing and location of Saturn's heavy-element accretion.}

\subsection{Robustness of the formation scenario}

{Our calculations describe the composition of an evolving PSN without including planetary accretion sinks, gap opening, pebble filtration, or the mutual dynamical evolution of Jupiter and Saturn. The inferred formation conditions therefore apply to material available near the \ce{CO2} iceline before the disk was substantially restructured by either planet. Because pebbles drift inward and Saturn formed exterior to Jupiter, Jupiter would not directly intercept the pebble flux before it crossed Saturn's orbit. However, once Jupiter reached pebble isolation mass, the resulting pressure maximum exterior to its orbit could halt inward pebble transport and modify the distributions of solids and vapors in the Jupiter--Saturn region \citep{Lambrechts2014,Lambrechts2014b,Bitsch2018}.}

{If Saturn acquired most of its heavy elements during the early $\sim$0.2--0.3 Myr interval favored here, before Jupiter substantially perturbed the disk, our compositional constraints remain applicable. If Jupiter instead reached pebble isolation earlier, its pressure bump could alter both the pebble reservoir and the vapor fluxes available near Saturn's formation region. Conversely, if Saturn reached pebble isolation first, it would reduce the pebble supply available to the growing Jupiter. Quantifying these interactions requires coupled models of simultaneous planetary growth, migration, and disk evolution and lies beyond the scope of the present work.}

{Our calculations also relate Saturn's inferred heavy-element inventory to the composition of the PSN at the time of formation. We compare our predictions with envelope metallicities inferred from present-day interior models but do not model Saturn's subsequent interior evolution. Modern interior models suggest that giant planets may undergo substantial post-formation redistribution of heavy elements through core erosion, the development of dilute cores, compositional mixing, and phase separation \citep[e.g.,][]{Militzer2019,Mankovich2021,Vazan2020,Knierim2025}. These processes may modify the present-day atmospheric composition and radial distribution of heavy elements without necessarily changing the total heavy-element inventory. Our inferred formation conditions should therefore be interpreted as constraints on the primordial material accreted by Saturn rather than on its present-day internal structure.}


\subsection{Model limitations and future work}

{Our calculations employ a one-dimensional viscous $\alpha$-disk model that captures the coupled transport of gas, solids, and volatile species
\citep{Aguichine2020,Aguichine2022} but neglects multidimensional gas flows, pressure bumps, zonal flows,
magnetically driven winds, and other disk substructures
\citep{Pinilla2012,Bai2016,Dullemond2018}. Likewise, we adopt a Schmidt number of unity
\citep{Armitage2020}, although different values would modify vapor diffusion rates and the strength of the cold-finger effect \citep{Yap2026}.}

{The adopted volatile abundances represent physically motivated initial conditions based on interstellar and Solar System measurements \citep{Boogert2015,LeRoy2015,Altwegg2019,Oberg2023},
rather than the outcome of a self-consistent thermochemical model \citep{Eistrup2016,Eistrup2018,vantHoff2020,Bosman2021}.}

{Finally, as discussed in Sect.~\ref{sec:formation-pt}, our simplified
pebble-to-planetesimal conversion prescription
\citep{Drazkowska2017} is intended only to estimate the amount of solid material available for planetary growth. It should not be interpreted as a self-consistent model of streaming instability or planetary core formation
\citep{Youdin2005,Johansen2007,Carrera2015,Yang2017}.}

{Despite these limitations, the results consistently identify material originating near the CO$_2$ iceline as the only solution capable of simultaneously reproducing Saturn's atmospheric volatile enrichments
and inferred envelope metallicity. Future models coupling multidimensional disk evolution, time-dependent chemistry, self-consistent planet formation, and long-term interior evolution will provide a more comprehensive picture of Saturn's origin.}

\section{Conclusion}\label{sec:ccls}

In this work, we investigated how icelines affected the composition of the PSN and whether this imprint could account for Saturn’s observed composition, while remaining agnostic to formation processes. We specifically examined conditions near the \ce{CO2} or CO iceline.

Our model predicts PSN metallicities consistent with those inferred for Saturn’s envelope (1–4.2 × protosolar), but not with the higher bulk metallicities. This suggests that iceline-driven enrichment in the PSN was insufficient to support Saturn’s formation via gravitational instability. Supersolar metallicities are primarily achieved interior to the \ce{H2O} iceline, largely independent of the $\alpha$ value; however, this region is excluded as it would correspond to Jupiter’s formation location. We further find that supersolar metallicities are only achieved near the \ce{CO2} iceline or interior to the \ce{H2O} iceline when $\alpha \le 10^{-3}$, and within 10 AU when $\alpha = 5\times10^{-3}$. By contrast, the PSN metallicity at the CO iceline is never supersolar, because this region is dynamically unfavorable for generating enrichment peaks. As a result, the CO iceline is not a viable formation location for Saturn under either the core accretion or gravitational instability scenarios. This is consistent with Saturn having formed well interior to the present-day orbits of the ice giants.

These observations led us to focus on the composition of the PSN at 6 AU, in the vicinity of the \ce{CO2} iceline, where strong C and O enrichments arise from the concentration of \ce{CO2} via the cold-finger effect. Following \cite{Aguichine2022}, we assumed that volatiles were delivered to Saturn through both the solid and vapor phases and examined whether Saturn’s atmospheric composition could have been inherited from the PSN. We found that enrichment peaks at 6 AU formed as early as $\sim$0.2–0.3 Myr for C and O, driven by the inward drift of \ce{CO2}-rich pebbles. By contrast, enrichment peaks for S, P, and noble gases developed at later times, reflecting their more distant icelines. Although higher values of $\alpha$ reduced the time required for vapors to diffuse from their icelines to 6 AU, they also diminished the amplitude of the resulting enrichment peaks.

{We performed a parameter study exploring plausible values of the turbulent viscosity parameter $\alpha$ and the uncertain partitioning of carbon between CO and \ce{CO2}. The largest C and O enrichments at 6 AU (up to $\sim$5 times protosolar) occur when carbon is predominantly in \ce{CO2} and $\alpha \le 10^{-3}$. However, no parameter combination reproduces the higher metallicities inferred for Saturn’s atmosphere (9--12.5 times protosolar).}

{Although our model adopts fixed volatile partitioning rather than coupling disk evolution to a time-dependent chemical network, these qualitative conclusions remain unchanged across the range of plausible CO:\ce{CO2} ratios explored in this study. Future work combining PSN evolution with self-consistent astrochemical models will be needed to quantify the effects of time-dependent volatile chemistry on Saturn's inferred formation conditions.}

{That said, the observed C enrichment could be reproduced at the \ce{CO2} iceline if all carbon resided in \ce{CO2}, with no refractory component. Our baseline assumes \textit{in situ} formation at this location. Higher metallicities may instead reflect radial migration over several AU, enhancing solid accretion \citep{Turrini21}, or the incorporation of carbon-rich solids scattered from Jupiter’s formation region \citep{Ronnet20}. We also find that colder disks yield higher enrichments: neglecting stellar irradiation increases the solid volatile fraction, strengthens inward pebble drift, and amplifies enrichment peaks. Consistently, \cite{Aguichine2022}, who modeled a non-irradiated disk, reported higher volatile enrichments than obtained here.}

The thermal structure of protoplanetary disks remains poorly constrained. Observations place the CO iceline between 48 and 75 AU and the \ce{N2} iceline between 78 and 145 AU in some disks, implying shallow thermal gradients \citep{Qi2019}, whereas other systems show CO icelines as close as 12–20 AU, indicative of much steeper gradients \citep{Oberg2021}. Steep gradients may result from disk shielding, either by the inner disk edge or by the accumulation of ice-rich pebbles at the water iceline \citep{Ohno2021}. In the latter case, regions just beyond the water iceline (beyond $\sim$4 AU) become cold enough for most volatiles to condense as pure ices, yielding Saturn compositions consistent with our non-irradiated disk models. This study considers only pure ices. In reality, volatiles may also be trapped in clathrate or amorphous ice \citep{Owen1999,Hersant2004,Mousis2012}. In the amorphous ice case, volatiles are released simultaneously during crystallization at $\sim$140 K, which has an effect of evening out all enrichments \citep{Mousis2019,Schneeberger2023}.

To place our results in the context of current planet-formation models \citep{Drazkowska2023}, we tested whether a simple prescription for planetesimal formation \citep{Drazkowska2017,Lichtenberg2021} applies to a disk containing multiple icelines. We found that planetesimal formation is not restricted to the water iceline but can occur at all icelines where the cold-finger effect produces sufficient solid pile-up. Up to half of the available trace species are converted into planetesimals within $\sim$0.2 Myr, comparable to the timescale for volatile enrichment at icelines. This suggests that pebble drift simultaneously enhances volatile abundances and triggers planetesimal formation by streaming instability at multiple icelines.

Planetesimals formed from local solids span a wide range of compositions, from solar (roughly equal rock and ice fractions) to pure rock, including water-rich bodies with negligible carbon and volatile-rich planetesimals dominated by \ce{CO2} near the \ce{CO2} iceline, with C/O ratio $\sim$0.5, which is very close to the protosolar value (C/O = 0.46). As a result, it may be difficult to distinguish whether Saturn accreted solar-composition solids or \ce{CO2}-dominated material based on its C/O ratio alone. Finally, the inclusion of planetesimals here demonstrates that pebble drift promotes conditions favorable to planetesimal formation and yields masses consistent with Saturn’s heavy-element inventory; determining Saturn’s final metallicity would require coupled gas and planetesimal accretion modeling, which lies beyond the scope of this work.

\begin{figure*}
	\centering
	\includegraphics[width=0.8\textwidth]{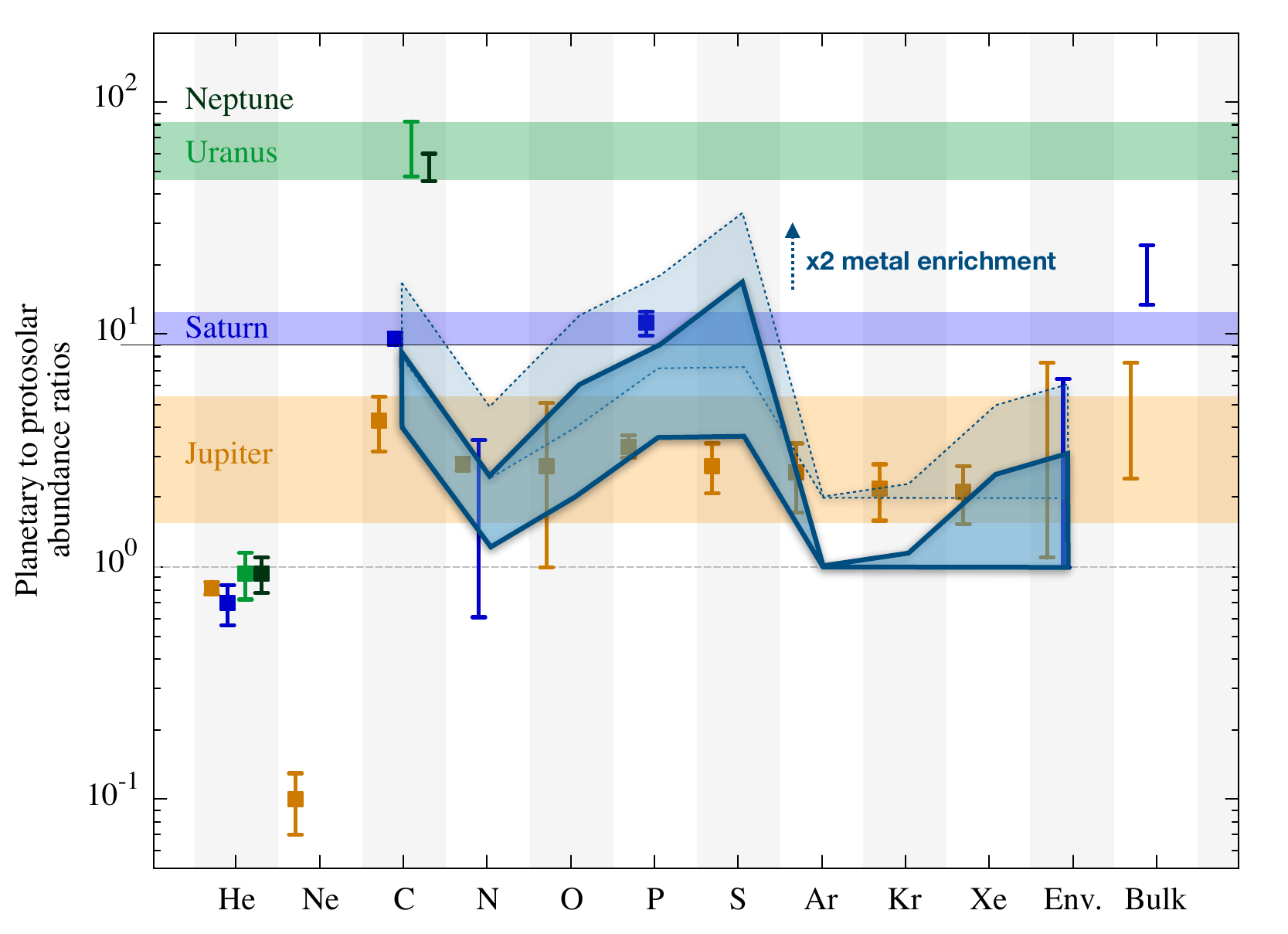}
	\caption{Summary of the predicted volatile abundances in Saturn’s atmosphere, {represented by the blue shaded region}. The dashed line shows the case where Saturn accretes twice as many metals, for instance due to less efficient gas accretion.}
	\label{fig:predictions}
\end{figure*}

The model predictions shown in Fig.~\ref{fig:predictions} are derived from disk compositions at the peak of carbon enrichment ($\sim0.3$ Myr) for CO:\ce{CO2} = 10:30, which yields the highest C enrichment at the \ce{CO2} iceline among all simulations (bottom row of Fig.~\ref{fig:enrichment-all}), including variants of the fiducial model (Fig.~\ref{fig:enrichment-fiducial-compare}). Noble gases serve as excellent indicators for the formation conditions of giant planets within the solar system \citep{Atreya2020}. An \textit{in situ} probe could definitively distinguish between different volatile sources in Saturn, ruling out multiple formation pathways \citep{Mousis2022_atmospheres}.

\section*{Acknowledgment}
This material is based upon work supported by NASA’S Interdisciplinary Consortia for Astrobiology Research (NNH19ZDA001N-ICAR) under award number 19-ICAR19\_2-0041. We acknowledge use of the lux supercomputer at UC Santa Cruz, funded by NSF MRI grant AST 1828315. A.A. thanks Jonathan Fortney and Naor Movshovitz for insightful comments that improved the clarity of this manuscript. We thank the anonymous reviewers for their insightful comments that helped improving the clarity of the manuscript.

\appendix
\section{Relationship Between Atmospheric Enrichments and Planetary Metallicity}
\label{app:metallicity}

{Atmospheric elemental abundances are commonly expressed as X/H ratios normalized to their protosolar values, whereas interior models characterize giant planets by their bulk or envelope metallicity. This appendix summarizes the relationship between these two quantities. The enrichment of an element X relative to its protosolar abundance is given by:}

\begin{equation}
	f_{\ce{X}} \equiv \left(\frac{\ce{X}}{\ce{H}}\right)^\ast = \frac{(\ce{X}/\ce{H})}{(\ce{X}/\ce{H})_{\odot}},\label{eq:def-enrichment}
\end{equation}

\noindent where the symbol $\ast$ denotes normalization to the protosolar value. In protoplanetary disk models, the metallicity $Z_{\ce{X}}$ of a heavy element $X$ is commonly defined as the ratio $Z_X = M_X/(M_{\ce{H}} + M_{\ce{He}})$, where $M_X$ is the mass of element $X$, and $M_{\ce{H}}$ and $M_{\ce{He}}$ denote the masses of hydrogen and helium, respectively.


As a mass ratio, $Z_{\ce{X}}$ differs from the mole ratio $(\ce{X/H})$. However, the two become equivalent when normalized to their respective protosolar values. Normalizing both quantities to their protosolar values gives:

\begin{equation}
	Z^\ast_{\ce{X}} \equiv \frac{(M_{\ce{X}}/M_{\mathrm{gas}})}{(M_{\ce{X}}/M_{\mathrm{gas}})_{\odot}}
	= f_{\ce{X}}.  \label{eq:def-metallicity}
\end{equation}

\noindent This equivalence allows atmospheric elemental enrichments to be compared directly with the metallicities predicted by disk models and inferred from giant-planet interior models.

The normalized bulk metallicity of a giant planet is:

\begin{equation}
	Z_{\mathrm{bulk}}^\ast = \frac{M_Z}{M_{\ce{H}}+M_{\ce{He}}} \cdot \frac{1}{Z_{\odot}} = \frac{M_Z}{M_{\mathrm{planet}}-M_Z} \cdot \frac{1}{Z_{\odot}},  \label{eq:metallicity-bulk}
\end{equation}

\noindent where $M_Z$ is the total mass of heavy elements in the planet (i.e., core and envelope), and $Z_{\odot}=0.0155$ (see below). Assuming the core consists entirely of heavy elements, the corresponding envelope metallicity is:

\begin{equation}
	Z_{\mathrm{env}}^\ast = \frac{M_{Z,\mathrm{env}}}{M_{\mathrm{planet}}-M_Z} \cdot \frac{1}{Z_{\odot}},  \label{eq:metallicity-env}
\end{equation}

\noindent where $M_{Z,\mathrm{env}}=M_Z-M_{\mathrm{core}}$. 


{Historically, giant-planet interiors were commonly represented by three-layer models consisting of a compact heavy-element core, an inner metallic hydrogen envelope, and an outer molecular hydrogen envelope, with the boundary between the two envelopes associated with the molecular-to-metallic hydrogen transition \citep{Stevenson1982,Saumon1995}. More recent interior models instead favor extended dilute cores and gradual heavy-element distributions rather than sharp compositional boundaries \citep{Wahl2017,Militzer2019,Mankovich2021}. Since our objective is to compare PSN compositions with the heavy-element inventory inferred from interior models, we use the envelope-averaged metallicity reported by these models, independent of the specific internal structure adopted.}

The metallicity $Z$ used here is not strictly equivalent to the PSN dust-to-gas ratio. {In the PSN, the dust-to-gas ratio represents only condensed solids, $Z_{\mathrm{solid}}$, whereas vaporized species contribute an additional metallicity $Z_{\mathrm{vapor}}$. The total metallicity is therefore $Z_{\mathrm{tot}} = Z_{\mathrm{solid}} + Z_{\mathrm{vapor}}$. Throughout this work, planetary heavy-element contents are expressed as the normalized mass ratio Z, related to the heavy-element mass fraction z through $Z = z/(1 - z)$. For the protosun, $z_{\odot} = 0.0153$, corresponding to $Z_{\odot} = 0.0155$ \citep{Lodders2009}.}
 
{These definitions are used throughout the paper when comparing PSN metallicities with the atmospheric enrichments and interior metallicities inferred for Saturn.}

\section{Adopted Disk Composition}
\label{app:disk}

{The PSN model assumes protosolar elemental abundances from \citet{Lodders2009}. Carbon, oxygen, nitrogen, sulfur, and phosphorus are partitioned among refractory and volatile reservoirs following the prescriptions described in Sec.~\ref{sec:methods}. The fiducial model adopts a CO:\ce{CO2} ratio of 10:30, while additional simulations explore CO:\ce{CO2} ratios of 30:10 and 40:00 to account for the observed diversity of volatile compositions in interstellar and Solar System materials.}

{Table~\ref{tab:compo-disk} summarizes the abundances of the trace species included in the calculations. The middle column lists abundances relative to hydrogen, and the right column gives the corresponding mass fractions relative to the H--He gas.}

\begin{table*}[ht]
\centering
\caption{{Initial abundances of the trace species included in the PSN model. The second column gives abundances relative to hydrogen, and the third column lists the corresponding mass fractions relative to the H--He gas. Ranges correspond to the different CO:\ce{CO2} ratios explored in this work.}}
\label{tab:compo-disk}

\begin{tabular}{lcc}
\hline
\hline
Trace species &
$(\mathrm{X}/\mathrm{H})_\odot$ &
$Z_k=\Sigma_{0,k}/\Sigma_{\mathrm g,0}$\\
\hline
\ce{H2O} & $(2.525-3.500)\times10^{-4}$ &
$(3.232-4.484)\times10^{-3}$\\
CO & $(0.325-1.300)\times10^{-4}$ &
$(0.647-2.590)\times10^{-3}$\\
CO$_2$ & $(0-9.749)\times10^{-5}$ &
$(0-3.052)\times10^{-3}$\\
CH$_4$ & $3.250\times10^{-6}$ &
$3.700\times10^{-5}$\\
CH$_3$OH & $5.427\times10^{-6}$ &
$1.236\times10^{-4}$\\
N$_2$ & $3.683\times10^{-5}$ &
$7.338\times10^{-4}$\\
NH$_3$ & $8.185\times10^{-6}$ &
$9.914\times10^{-5}$\\
H$_2$S & $8.165\times10^{-6}$ &
$1.978\times10^{-4}$\\
PH$_3$ & $3.184\times10^{-7}$ &
$7.699\times10^{-6}$\\
Ar & $3.573\times10^{-6}$ &
$1.015\times10^{-4}$\\
Kr & $2.153\times10^{-9}$ &
$1.238\times10^{-7}$\\
Xe & $2.104\times10^{-10}$ &
$1.964\times10^{-8}$\\
\hline
Total volatiles & -- &
$(7.548-8.374)\times10^{-3}$\\
Total refractory material & -- &
$(7.126-7.952)\times10^{-3}$\\
\hline
\end{tabular}
\end{table*}

\bibliographystyle{aasjournal}
\bibliography{sn-bibliography}



\end{document}